# Transition between heavy-fermion strange metal and quantum spin liquid in a 4d-electron trimer lattice

Hengdi Zhao[1], Yu Zhang[1], Pedro Schlottmann[2], Rahul Nandkishore[1,3] Lance E. DeLong[4], and Gang Cao[1,5*]

[1]*Department of Physics, University of Colorado at Boulder, Boulder, CO 80309, USA*

[2]*Department of Physics, Florida State University, Tallahassee, FL 32306, USA*

[3]*Center for Theory of Quantum Matter, University of Colorado at Boulder, Boulder, CO 80309, USA*

[4]*Department of Physics and Astronomy, University of Kentucky, Lexington, KY 40506, USA*

[5]*Center for Experiments on Quantum Materials, University of Colorado at Boulder, Boulder, CO 80309, USA*

We present experimental evidence that a heavy Fermi surface consisting of itinerant, charge-neutral spinons underpins both heavy-fermion-strange-metal (without f electrons) and quantum-spin-liquid states in the 4d-electron trimer lattice, $Ba_4Nb_{1-x}Ru_{3+x}O_{12}$ ($|x| < 0.20$). These two exotic states both exhibit an extraordinarily large entropy, a linear heat capacity extending into the milli-Kelvin regime, a linear thermal conductivity at low temperatures, and separation of charges and spins. Furthermore, the insulating spin liquid is a much better thermal conductor than the heavy-fermion-strange-metal that separately is observed to strongly violate the Wiedemann-Franz law. We propose that at the heart of this 4d system is a universal, heavy spinon Fermi surface that provides a unified framework for explaining the exotic phenomena observed throughout the entire series. The control of such exotic ground states provided by variable Nb concentration offers a new paradigm for studies of correlated quantum matter.

*Corresponding author. Email: gang.cao@colorado.edu

Planckian strange metals **[1-15]**, heavy-fermion metals **[16-26]** and quantum spin liquids **[27-34]** are intriguing quantum states of matter currently subject to intense investigation. *Planckian metals* feature a linear temperature T dependence of electrical resistivity ρ that persists to T = 0 K, independent of the Fermi surface topology, because charge carriers scatter at a rate that saturates in the Planckian limit, $1/\tau = \alpha k_B T/\hbar$ (where τ is the relaxation time, $k_B$ and $\hbar$ the Boltzmann and reduced Planck's constants, respectively, and α a constant of order unity **[2-11]**).

*Heavy-fermion metals*, on the other hand, are characterized by a large Sommerfeld coefficient γ arising from electronic contributions to low-T heat capacity C(T). Values of γ range from 20 mJ/mole $K^2$ **[17]** to up to 1600 mJ/mole $K^2$ **[19]**. At the present time, the heavy-fermion behavior is observed almost exclusively in 4f- and 5f-electron materials and is dictated by hybridization between itinerant and localized magnetic electron states **[21-25]**. A notable exception to the ongoing predominance of heavy f-electron states is the transition metal oxide $LiV_2O_4$. This material exhibits a very high γ = 420 mJ/mole $K^2$ **[35-37]** and $\rho \propto T^2$ in the low-T limit **[37]**. The heavy-fermion behavior is thought to be a consequence of its frustrated lattice **[38-40]**. Heavy-fermion fractionalization driven by an interplay between geometric frustration and Kondo effect also has been extensively discussed in recent years in part because the one-dimensional Doniach "Kondo necklace" scenario **[41]** may not be flexible enough to account for non-Fermi liquid behavior in heavy-fermion systems **[e.g.,18, 21, 25, 26]**.

Finally, *quantum spin liquids* are among the most intensively studied states of condensed matter. Theoretical treatments of the honeycomb iridates, α-$RuCl_3$, and other triangular lattices have inspired a large body of experimental work that seeks to identify various types of quantum spin liquids whose experimental signatures include a linear C(T) extending to T = 0 K, which is expected to arise from a Fermi surface of charge-neutral spinons **[27-34, 42]**.

Each of the three states mentioned above typically occurs in different classes of materials with distinct energy scales. In contrast, we have synthesized the single-crystal series $Ba_4Nb_{1-x}Ru_{3+x}O_{12}$ ($|x| < 0.20$) (**Fig. 1a-1c**) that, depending on composition x, exhibits the characteristics of all three exotic states, as well as completely novel phenomena. At one end of the series, a novel metal *simultaneously* exhibits the characteristics of both a *strange* metal ($\alpha$ is of order unity; see **Fig. 1d**) and a *heavy-fermion* metal ($\gamma$ = 164 mJ/mole $K^2$; see **Fig.1e**). Furthermore, this metal violates the Wiedemann-Franz (WF) law by up to a factor of 26 (**Fig. 1f**) and exhibits a sign-change in the Hall effect when the current is applied along different crystalline directions (**Fig.S6**). We term this material a *heavy-fermion strange metal*. At the other end of the series, a strongly *frustrated Mott insulator* exhibits an increased $\gamma$ = 225 mJ/mole $K^2$ and an enhanced thermal conductivity $\kappa$ greater than that of the metal. A *heavy-fermion metal* with $\gamma$ = 181 mJ/mole $K^2$ is situated between these two end states (**Fig. 1h)**. The electrical resistivity $\rho$ progressively changes with Nb content (**Fig. 1g**), but a strikingly linear C(T) extending down to 50 mK, an equally linear low-T $\kappa$ and a paramagnetic state with a strong exchange energy of up to 340 K persist throughout the entire series (**Figs. 1h-1i**). A large entropy with no discernible long-range order extends into the milli-Kelvin regime for all Nb compositions studied (**Fig.1h**).

The larger $\kappa$ of the insulating state compared to that of the metallic state forcefully argues that a heavy Fermi surface of *itinerant charge-neutral spinons* must be invoked to explain this behavior throughout the series (**Fig. 1h-1i**). Importantly, it rules out the possibility that the large linear-T C(T) in the insulator comes from localized degrees of freedom that could not contribute to $\kappa$. We stress that this scenario explains a large array of exotic phenomena in terms of a dissociation of charges and spins, as discussed below. This rich, novel phenomenology arises from a rare interplay between geometric frustration and competing interactions and provides a compelling new paradigm for correlated quantum materials. Note that polycrystalline $Ba_4NbRu_3O_{12}$ (x = 0) has been reported to be a geometrically frustrated insulator with $\gamma$ = 31 mJ/mole $K^2$ with spin freezing near 4 K **[43]**, which is not discerned in this study.

Our single-crystal x-ray diffraction data gathered between 100 K and 300 K indicate that the series of $Ba_4Nb_{1-x}Ru_{3+x}O_{12}$ (for |x| < 0.20; the sign of x can be either positive or negative) adopts a rhombohedral structure with the space group R-3 (No. 148), which is retained for all x studied herein (**Fig. S1** and **Tables S1-S3 [44]**). For simplicity, we use $Nb_{0.81}$, $Nb_{0.95}$, $Nb_{1.00}$ and $Nb_{1.16}$ to denote $Ba_4Nb_{1-x}Ru_{3+x}O_{12}$ with different x-values. The crystal structural data indicate that the formal valence of Nb is 5+ ($4d^0$) across the entire series **[44]**. The average valence of Ru decreases with increasing Nb content from 3.75+ for $Nb_{0.81}$, to 3.69+ for $Nb_{0.95}$ and to 3.59+ for $Nb_{1.16}$, indicating a mixed valence of $Ru^{3+}$($4d^5$) and $Ru^{4+}$ ($4d^4$) which is common among perovskite ruthenates **[45]**. The average number of 4d electrons in each $Ru_3O_{12}$ trimer is estimated to be 12.75, 12.93, 13 and 13.23 in $Nb_{0.81}$, $Nb_{0.95}$, $Nb_{1.00}$, and $Nb_{1.16}$, respectively.

The key structural element of $Ba_4Nb_{1-x}Ru_{3+x}O_{12}$ is the $Ru_3O_{12}$ trimer formed by three face-sharing $RuO_6$ octahedra. The trimers (red) are linked along the *c* axis (**Fig. 1a**) by corner-sharing $NbO_6$ octahedra (blue) to form the triangular *ab* planes or trimer layers, which is a common source of geometric frustration (**Figs. 1b, 1i**) **[44]**.

We discuss below the definitive characteristics of each of the three exotic states shown in **Fig. 1i**. Additional data are included in the Supplemental Material **[44]**.

***Heavy-fermion strange metal***

The magnetic susceptibility $\chi$ of $Nb_{0.81}$ exhibits a robust Curie-Weiss behavior down to 1.7 K (**Fig. 2a)**. An analysis of the data for 50-350 K yields the Curie-Weiss temperature $\theta_{CW}$ = -169 K and -128 K for the *a*-axis $\chi_a$ and *c*-axis $\chi_c$, respectively, which reflects a strong antiferromagnetic coupling. The paramagnetic susceptibility is on the order $10^{-3}$ emu/mole, comparable to the values observed for heavy-fermion systems **[19]** (See **Table S4**).

The resistivities $\rho$ of $Nb_{0.81}$ from 50 mK to 380 K for current along the *a*-axis $\rho_a$ and *c*-axis $\rho_c$ are shown in **Figs. 2b-2d** (note that $\rho_a > \rho_c$). The linearity in T in both $\rho_a$ and $\rho_c$ for T > 250 K is a common occurrence in many correlated oxides due to enhanced electron-electron and electron-phonon interactions **[15]**. What is intriguing is that the linearity in both $\rho_a$ (= $A_aT$) and $\rho_c$ (= $A_cT$) persists from 50 K down to 50 mK (**Figs. 1d, 2c, 2d**), with slope values $A_a = 2.45 \times 10^{-6}$ $\Omega$ cm/K and $A_c = 1.88 \times 10^{-6}$ $\Omega$ cm/K. Our estimate of the Planckian limit yields lower bounds for $\alpha = 0.46$ and 0.75 for the *a* axis and *c* axis, respectively **[44]**, confirming that the scattering rate of charge carriers in $Nb_{0.81}$ is indeed close to the Planckian limit (**Fig. S2a**). Note that the Hall coefficient changes sign with current orientation, indicating that the Fermi surface of the metallic $Nb_{0.81}$ is highly anisotropic (**Fig.S6**) **[44]**.

The linearity of $\rho_a$ and $\rho_c$ persists in the presence of applied pressure P up to 2.4 GPa (right scale in **Figs. 2c-2d**), reflecting that the scattering rate is insensitive to any possible changes in the crystal and/or electronic structures due to applied pressure. However, application of H changes scattering processes, as indicated by a nearly $T^2$-dependence for both $\rho_a$ and $\rho_c$ (**Figs. 2c-2d**, **S2b**).

As for the low-T heat capacity C(T) at 14 T (**Fig. 2e)**, the slope of the linear T-dependence of C(T) changes only slightly from $\gamma = 164$ mJ/mole $K^2$ at H = 0 to $\gamma = 158$ mJ/mole $K^2$ at $\mu_oH = 14$ T (**Fig. 2e**). It is particularly striking that the high $\gamma$ value persists in such a strong field that normally depresses entropy, but this behavior is consistent with C(T) being dominated by a spinon Fermi surface **[27, 28, 34, 42, 46]**.

It is curious that the $T^3$-contribution to C(T) below 7 K (parameterized by the second term in $C(T) = \gamma T + \beta T^3$) is essentially zero (**Fig.2f**). The phonon contribution (positive $\beta$) is compensated by the second term of the Sommerfeld expansion of the electronic contribution (negative $\beta$) yielding a measured $\beta$ that is essentially zero. The low-T C(T) is then $\gamma T$ (**Fig. 2f**) **[44]**. Note that $\gamma$ (= C/T) rapidly rises to 275 mJ/mole $K^2$ at 50 mK. The nearly 70% increase in $\gamma$ is intriguing since $\gamma$ is not expected to be a strong function of T in the low-T limit in the absence of magnetic order, but it hints at a low-energy scale ~ 3 K. Note that this upturn in C/T is supplanted by a downturn in $Nb_{1.16}$, as discussed below, which helps rule out a nuclear Schottky anomaly with a signature term in C(T) ~ $1/T^2$ (**Fig. S3b**); this is also confirmed by our heat capacity data on $BaRuO_3$ and $Nb_2O_5$ measured in 50 mK – 1.0 K and at $\mu_oH$ = 0 and 14T, which do not follow at all C(T) ~ $1/T^2$, indicating a clear absence of any discernible nuclear Schottky contributions to C(T) of $Nb_{0.18}$ **[44, 47]**.

The thermal conductivity $\kappa_a(0)$ of $Nb_{0.81}$ along the *a* axis also decreases *linearly* with decreasing T below 10 K (**Fig. 3a**). At $\mu_oH_{||c}$ = 14 T, $\kappa_a$ responds strongly to H for 2 < T < 60 K, but very weakly beyond this range. The difference $\kappa_a(0)$ - $\kappa_a(14T)$ yields $\Delta\kappa_a$ that reflects contributions from heat carriers that are susceptible to magnetic fields (**Fig. 3a**) **[44]**. The same is true for $\kappa_c$ (**Fig. S4**) **[44]**.

Thermal and electrical currents are normally carried by the same quasiparticles. This is reflected in the WF law, $\kappa/\sigma = L_oT$, which states that for a single band the ratio of $\kappa$ to electrical conductivity $\sigma$ is proportional to T with a constant of proportionality = Lorenz number $L_o$. Here, using $\Delta\kappa_a$, $L(T) = \Delta\kappa_a\rho_a/T$ ($\sigma \cong 1/\rho$) we obtain $L > L_o$ by a factor of up to 26 (see squares in **Fig. 3b** and **Fig. 1f**). Such a strong violation of the WF law suggests that the relaxation times are vastly different for thermal and electrical processes or that more than one band is participating.

The WF law works well for typical metals, including Fermi liquids **[49]** when the condition of elastic scattering is satisfied **[50]**. Inelastic processes result in different relaxation times for charge and heat transport **[44]**. The strong violation of the WF law is likely due to a combined effect of geometric frustration and low dimensionality that causes a dissociation of charges and spins **[51-54]**, whose presence is reinforced by the inverse relationship between $\kappa_a$ and $\sigma_a$. The electrical insulator $Nb_{1.16}$, whose $\rho_a$ is $10^7$ greater than that for the metal $Nb_{0.81}$ at low T, is a much better thermal conductor than the metal $Nb_{0.81}$ below 30 K (**Fig. 3c**)! This is clearer in **Fig. 3d** in which the intercept of $\Delta\kappa_a/T$ at T = 0 for *the insulating* $Nb_{1.16}$ *is nearly twice as large as that for the metallic* $Nb_{0.81}$. Note the linearity of $\kappa_a(0)$ for both $Nb_{0.81}$ and $Nb_{1.16}$ below 10 K (**Fig. 3c**).

***Heavy fermion metal and frustrated insulator***

Increasing Nb content in $Ba_4Nb_{1-x}Ru_{3+x}O_{12}$ weakens the metallic state and eventually induces the insulating state. The weakened metallic state in $Nb_{0.95}$ and the insulating state in $Nb_{1.16}$ retain paramagnetic spin correlations (**Figs. 4a-4b**) with a strong exchange energy (absolute values of $\theta_{CW}$) up to 340 K and enhanced $\gamma$ values (**Fig. 1h**). In $Nb_{0.95}$, both $\rho_a$ and $\rho_c$ decrease below 150 K (**Fig. 4c**). The insulator $Nb_{1.16}$ displays a drastic increase in $\rho$ by more than 7 orders of magnitude with decreasing T (**Fig. 4d**). However, *this electrical insulator* $Nb_{1.16}$ *is a much better thermal conductor than the metal* $Nb_{0.81}$ (**Figs. 3c-3d)**. This contrasting behavior further implicates the dissociation of charges and spins and itinerant spinons (more data and discussion on $\kappa$ at magnetic fields are presented in **Fig.S7** and **[44])**.

C(T) of both $Nb_{0.95}$ and $Nb_{1.16}$ remains essentially linear (with slight deviations) below 0.5 K and yields even larger values of $\gamma$ = 181 and 225 mJ/mole K$^2$ for $Nb_{0.95}$ and $Nb_{1.16}$, respectively (**Fig. 4e**). The linearity of C(T) in both $Nb_{0.95}$ and $Nb_{1.16}$ also remains unchanged at 14 T and higher temperatures, ruling out the influences of magnetic impurities and further emphasizing the persistent Fermi surface of charge-neutral spinons (**Fig. S5**). Note that the linearity of C(T) in the insulating phase cannot be due to localized excitations since such excitations would not contribute to $\kappa$. The abrupt upturn in C/T vs $T^2$ below 0.5 K in $Nb_{0.81}$ is replaced by an equally abrupt downturn in $Nb_{1.16}$ (**Fig. 4f**) **[44]**.

*The most striking feature shared by the entire $Ba_4Nb_{1-x}Ru_{3+x}O_{12}$ series is the persistent linearity in low-T C(T) and $\kappa$, and extraordinarily large $\gamma$, independent of the ground state type (**Figs. 3a, 3c, and 4e-4f**)*. All our results constitute compelling evidence for the existence of a heavy-fermion strange metal (**Figs. 1d-1e**) **[44]** and a quantum spin liquid (**Figs. 3c-3d, 4d-4f**) depending on Nb content. The complex behavior of $Ba_4Nb_{1-x}Ru_{3+x}O_{12}$ suggests a powerful simplifying principle: At the heart of this 4d system is *a universal, itinerant, heavy spinon Fermi surface* that provides a unified framework for explaining the exotic phenomena observed throughout the entire series:

**(1)** The large exchange energy ($\theta_{CW}$) but no magnetic order (**Figs. 1h**);

**(2)** The linear C(T) persistent down to 50 mK, indicating a non-zero low-energy density of states (**Figs. 2e, 4e**);

**(3)** The extraordinarily large $\gamma$ only seen in heavy-fermion materials (**Figs. 1h, 4e**);

**(4)** The linear $\kappa$ at low T in all states, and much larger $\kappa$ in the insulating spin liquid than in the heavy-fermion strange metal (**Figs. 3c-3d**), which separately is observed to strongly violate the Wiedemann-Franz law (**Figs. 1d-1f)**.

In summary, *$Ba_4Nb_{1-x}Ru_{3+x}O_{12}$ is a rare system, if not the first, to encompass a proximity or even coexistence of both spin liquid physics, strong electron correlations tunable across a Mott metal-to-insulator transition, and a heavy-fermion strange metal.* Clearly, the rich, exotic phenomenology offers a compelling new paradigm of correlated quantum matter.

**Acknowledgements** G.C. thanks Minhyea Lee, Feng Ye, Daniel Dessau, and Longji Cui for useful discussions. This work is supported by National Science Foundation via Grant No. DMR 2204811.

Work by R.N. was supported by the U.S. Department of Energy (DOE), Office of Science, Basic Energy Sciences (BES) under Award # DE-SC0021346.

**Figure captions**

**Fig. 1. Key features and phase diagram. a,** The crystal structure of $Ba_4Nb_{1-x}Ru_{3+x}O_{12}$ along the *c* axis, and **b,** the *ab* plane. **c,** A single crystal of $Nb_{0.81}$. The T-dependence for 50 mK ≤ T ≤ 10 K for $Nb_{0.81}$, **d**, the *a*-axis $\rho_a$ and the *c*-axis $\rho_c$ (black dashed lines are linear fit) and, **e,** C(T). **f,** The Lorenz number of the WF law $L_o = \kappa/\sigma T$ and the measured $\kappa_a\rho_a/T$ for 1.8 ≤ T ≤ 20 K for $Nb_{0.81}$. **g,** the T-dependence of $\rho_a$ for 50 mK < T < 380 K for $Nb_{0.81}$, $Nb_{0.82}$, $Nb_{0.95}$, and $Nb_{1.16}$. **h,** $\gamma$ (blue) and $\theta_{CW}$ (red, right scale) as a function of Nb content. **i,** A schematic phase diagram.

**Fig. 2. Physical properties of $Nb_{0.81}$.** The T-dependence of **a,** the *a*-axis and *c*-axis $\chi_a$ and $\chi_c$ along with $\Delta\chi_a^{-1}$ and $\Delta\chi_c^{-1}$ (right scale) and **b,** $\rho_a$ and $\rho_c$ for 50 mK ≤ T ≤ 380 K. **c-d,** $\rho_a$ and $\rho_c$ for 50 mK < T ≤ 40 K at $\mu_oH_{||c} = 0$ and 14 T, and at P = 2.4 GPa and H = 0 (right scale) (black dashed lines are linear fit). **e,** The T-dependence of C(T) at $\mu_oH_{||c} = 0$ and 14 T. **f,** C/T vs. $T^2$.

**Fig. 3. Thermal conductivity of $Nb_{0.81}$ and $Nb_{1.16}$.** The T-dependence of **a,** $\kappa_a(0)$ at H = 0, $\kappa_a(14T)$ and $\Delta\kappa_a = \kappa_a(0) - \kappa_a(14T)$ below 60 K, and **b,** $L(T)/L_o$ where $L_o = 2.45 \times 10^{-8}$ ($V^2/K^2$) and the measured $L(T) = \Delta\kappa_a\rho_a/T$. Inset: the configuration for $\kappa_a$ measurements; $J_h$ is thermal current density. **c,** The T-dependence of $\kappa_a$ and $\rho_a$ (right scale) for $Nb_{0.81}$ and $Nb_{1.16}$. Note the inverse relationship between $\kappa_a$ and $\rho_a$ for $Nb_{1.16}$. **d,** $\Delta\kappa_a/T$ vs $T^2$ for metallic $Nb_{0.81}$ and insulating $Nb_{1.16}$. Note the intercept of $\Delta\kappa_a/T$ at T = 0 K is larger for $Nb_{1.16}$ than for $Nb_{0.81}$.

**Fig. 4. Physical properties of $Nb_{0.95}$ and $Nb_{1.16}$.** The T-dependence of $\chi_a$ and $\chi_c$ along with $\Delta\chi_a^{-1}$ and $\Delta\chi_c^{-1}$ (right scale) for **a,** $Nb_{0.95}$ and **b,** $Nb_{1.16}$. The T-dependence of $\rho_a$ and $\rho_c$ for **c**, $Nb_{0.95}$ and **d,** $Nb_{1.16}$. **e**, C(T) for 50 mK ≤ T ≤ 7 K and **f,** C/T vs $T^2$ for $Nb_{0.95}$, $Nb_{1.16}$ and $Nb_{0.81}$.

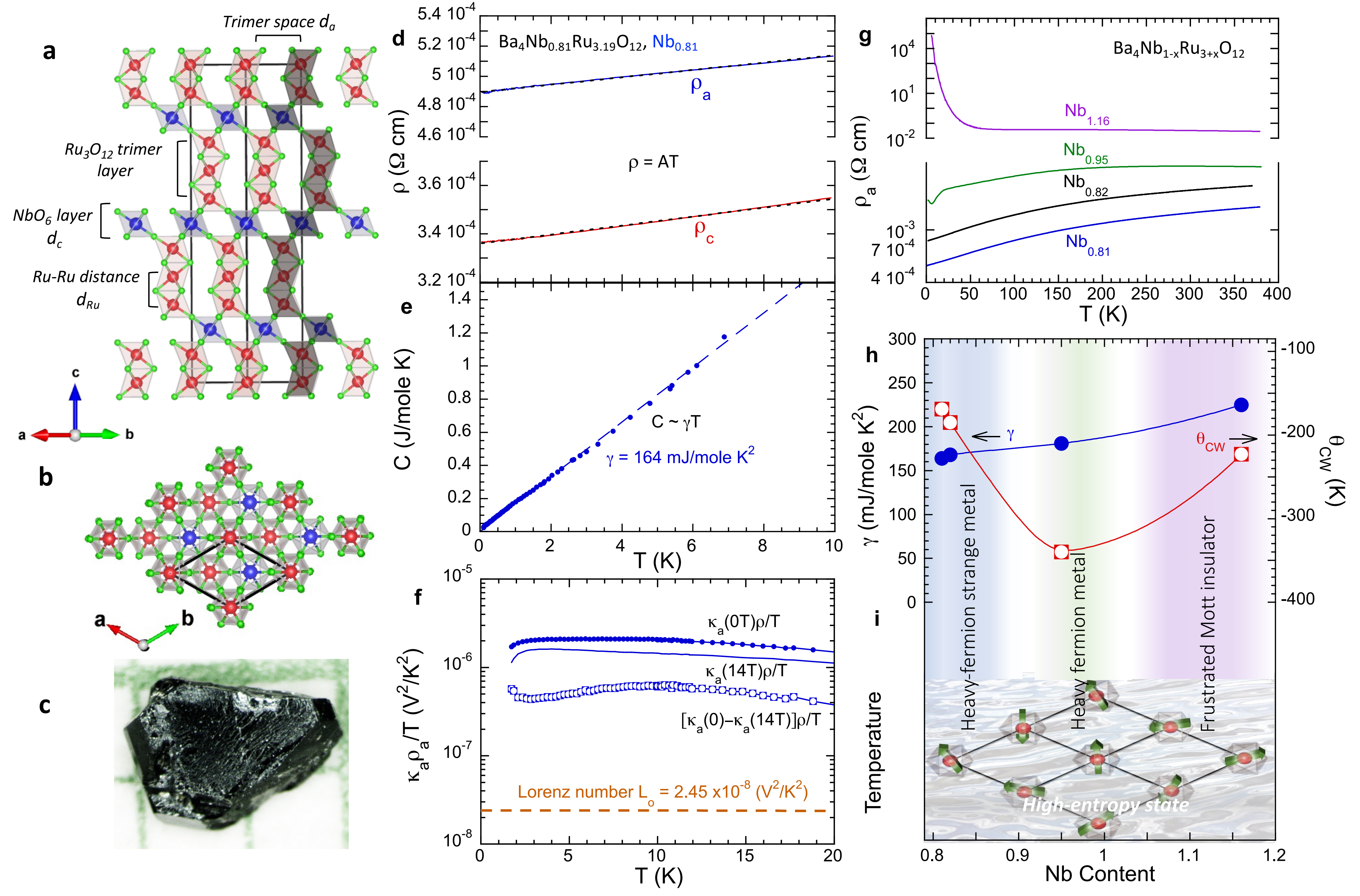

Figure 1

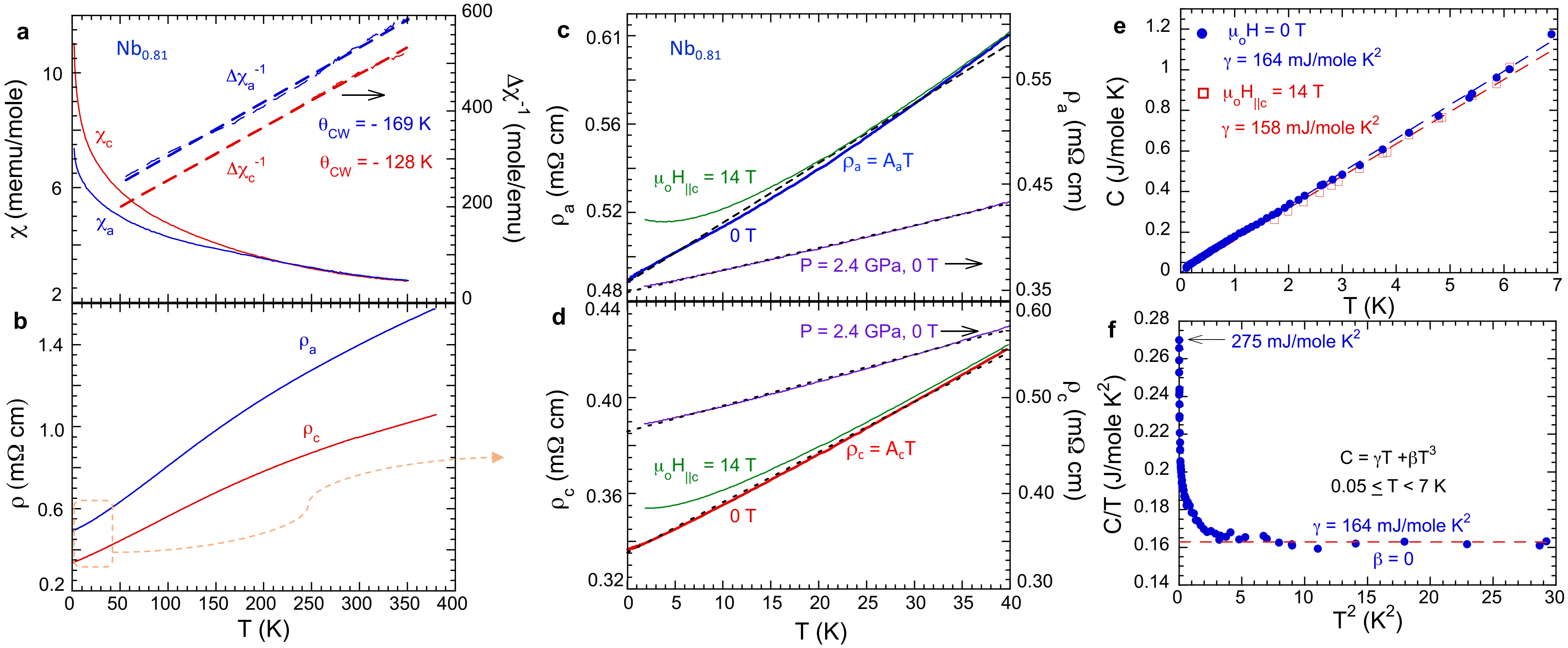
a
Nb0.81
Δχa-1
Δχc-1
θCW = - 169 K
θCW = - 128 K
χc
χa
χ (memu/mole)
Δχ-1 (mole/emu)
b
ρa
ρc
ρ (mΩ cm)
T (K)
c
Nb0.81
μoH||c = 14 T
ρa = AaT
0 T
P = 2.4 GPa, 0 T
ρa (mΩ cm)
d
P = 2.4 GPa, 0 T
μoH||c = 14 T
ρc = AcT
0 T
ρc (mΩ cm)
T (K)
e
μoH = 0 T
γ = 164 mJ/mole K2
μoH||c = 14 T
γ = 158 mJ/mole K2
C (J/mole K)
T (K)
f
275 mJ/mole K2
C = γT +βT3
0.05 ≤ T < 7 K
γ = 164 mJ/mole K2
β = 0
C/T (J/mole K2)
T2 (K2)

Figure 2

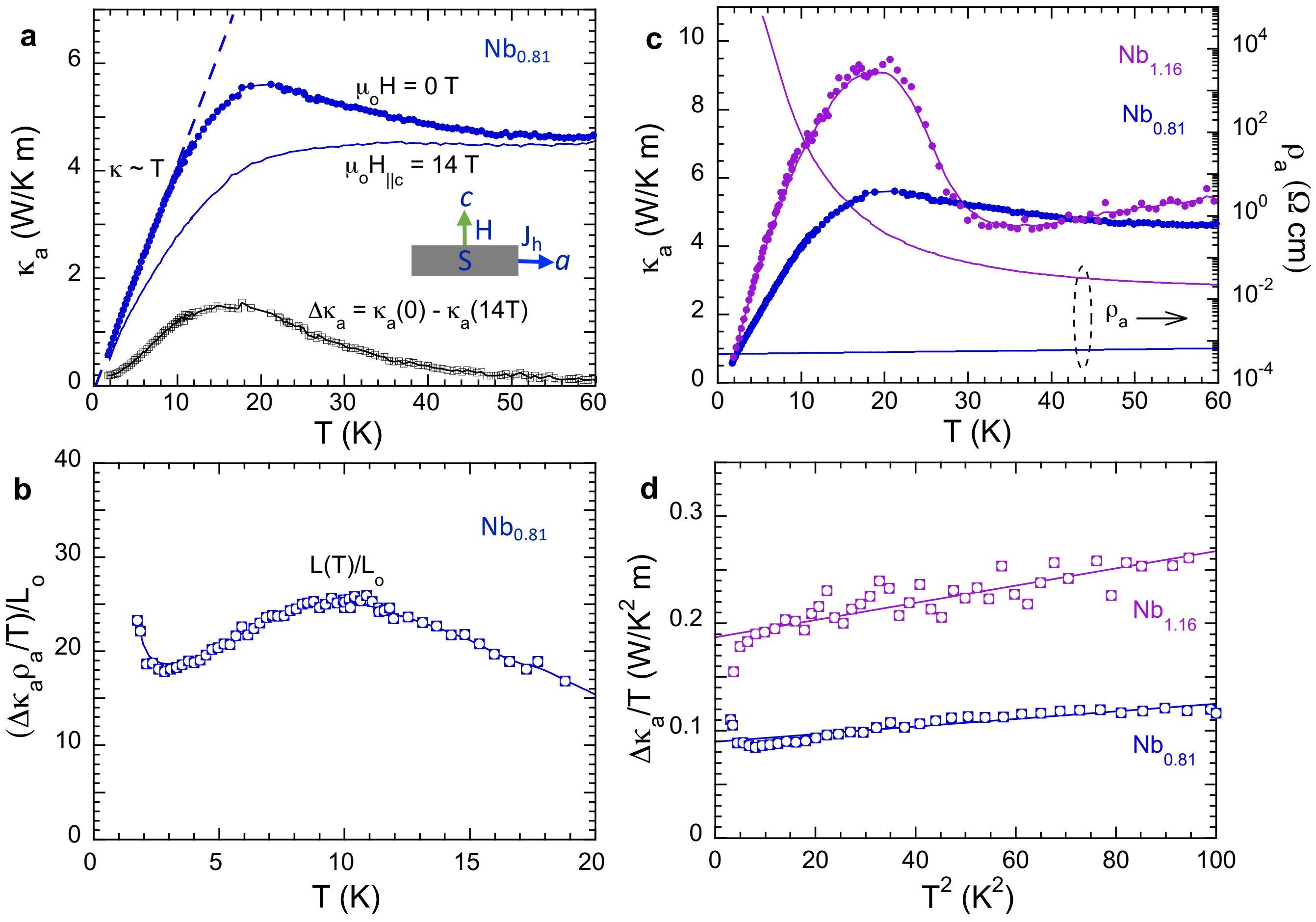

Figure 3

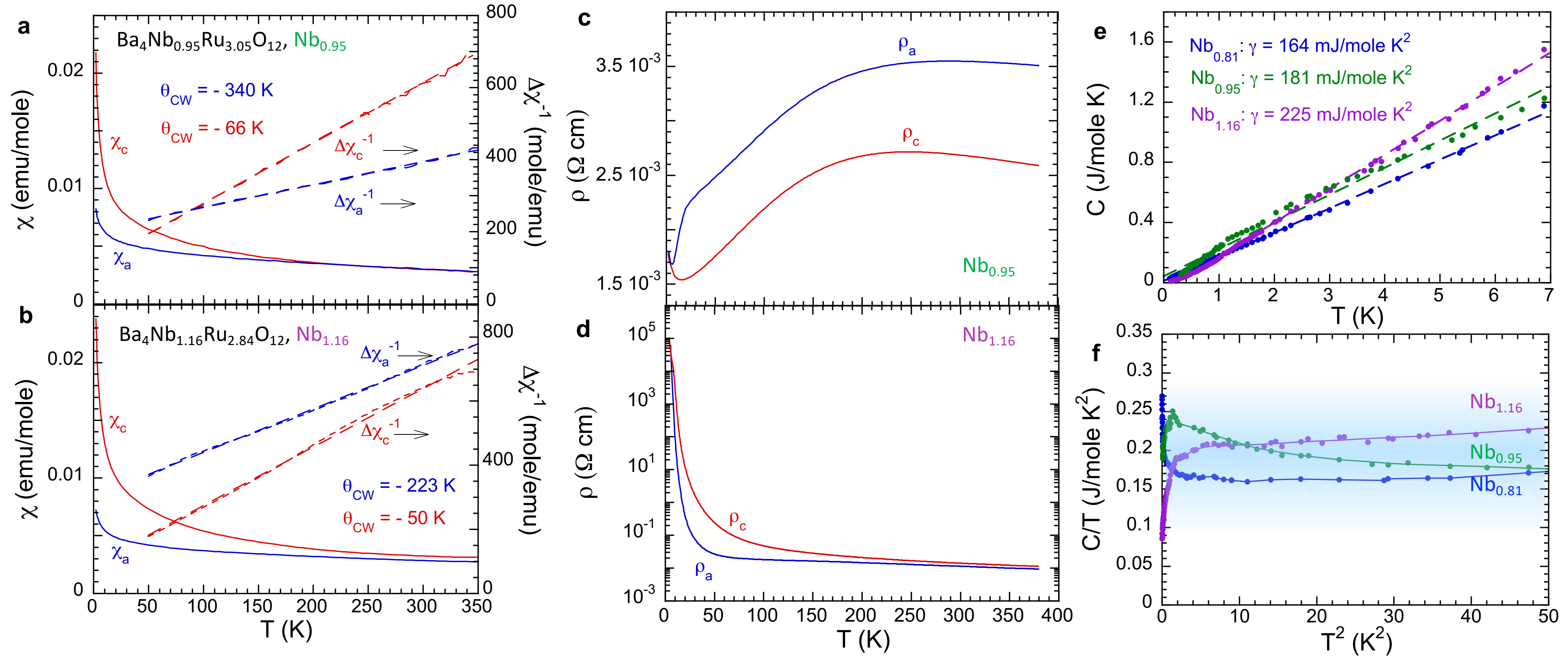

a
$Ba_4Nb_{0.95}Ru_{3.05}O_{12}$, $Nb_{0.95}$
$\theta_{CW}$ = - 340 K
$\theta_{CW}$ = - 66 K
$\chi_c$
$\chi_a$
$\Delta\chi_c^{-1}$
$\Delta\chi_a^{-1}$
χ (emu/mole)
$\Delta\chi^{-1}$ (mole/emu)
b
$Ba_4Nb_{1.16}Ru_{2.84}O_{12}$, $Nb_{1.16}$
$\theta_{CW}$ = - 223 K
$\theta_{CW}$ = - 50 K
T (K)
c
$\rho_a$
$\rho_c$
$Nb_{0.95}$
ρ (Ω cm)
d
$Nb_{1.16}$
e
$Nb_{0.81}$: γ = 164 mJ/mole $K^2$
$Nb_{0.95}$: γ = 181 mJ/mole $K^2$
$Nb_{1.16}$: γ = 225 mJ/mole $K^2$
C (J/mole K)
f
$Nb_{1.16}$
$Nb_{0.95}$
$Nb_{0.81}$
C/T (J/mole $K^2$)
$T^2$ ($K^2$)

Figure 4